\documentclass[journal=nalefd,manuscript=article,layout=twocolumn]{achemso}
\usepackage{amsmath}
\usepackage{amssymb}
\usepackage{color}
\author{Masaki Noro}
\affiliation[Tokyo Institute of Technology]
{Department of Physics, Tokyo Institute of Technology, Tokyo, Japan}
\author{Junya Tanaka}
\author{Takehito Yokoyama}
\affiliation[Tokyo Institute of Technology]
{Department of Physics, Tokyo Institute of Technology, Tokyo, Japan}
\author{Shuichi Murakami}
\email{murakami@stat.phys.titech.ac.jp}
\affiliation[Tokyo Institute of Technology]
{Department of Physics, Tokyo Institute of Technology, Tokyo, Japan}
\alsoaffiliation[TIES]
{TIES, Tokyo Institute of Technology, Tokyo, Japan}
\title[Theory of Chiral Transport in Carbon Nanotubes]
  {Theory of Chiral Transport in Carbon Nanotubes}

\begin{document}





\begin{abstract}
Based on the similarity between the chiral nanotube and the classical solenoid,
we study chiral transport along the circumferential direction in a
carbon nanotube. We calculate the chiral conductivity, representing a circumferential current induced by an electric field 
along the nanotube axis for various chiralities of carbon nanotubes. 
We find that the chiral conductivity in a chiral nanotube is in general non zero, and 
oscillates as a function of the Fermi energy. This oscillating behavior is 
attributed to the subband structure of the nanotubes and the warping of the 
Fermi surfaces. 
\end{abstract}


\vspace{1cm}

\section{Introduction}
Carbon nanotubes \cite{Iijima_1991, Iijima_1993} have been attracting much 
interest for their unique properties and for their possibilities of applications for electronic devices, 
such as
a field-emission display \cite{Liu_2010} and nanocomposite materials 
\cite{Byrne_2010}, nanosensors \cite{Allen_2007}.
Its novel transport properties has been revealed experimentally. For example, quantized conductance is measured \cite{Kong_2001}, 
and the ballistic conduction for both single \cite{Javey_2003} and multi-wall\cite{Berger_2002} 
carbon nanotubes are reported. 
Theoretically, a carbon nanotube can be either a metal or a semiconductor, 
depending on its chirality. This property is understood from the band structure 
of a graphene and periodic boundary conditions along the circumference direction. 
Because of the singular band structure and unique structure of a carbon nanotube, its transport properties are expected to be intriguing 
\cite{Miyamoto_1994,Ando_2000,Ando_1998,Nakanishi_1999}. 
 
In a chiral carbon nanotube, lower crystallographic symmetry allows chiral transport, i.e.
a current in the circumferential direction induced by an electric field along the tube. 
Namely, when carriers are doped, an electric current along the tube axis induces 
a current around the tube. In a previous study, the chiral transport along the circumferential 
direction has been suggested in a BC$_2$N nanotube theoretically \cite{Miyamoto_1994}, 
although the BC$_2$N nanotube has not been fabricated yet 
\cite{Ayala_2010}. In Ref.~\cite{Miyamoto_1994}, both first-principles and 
tight-binding calculations are performed for the BC$_2$N nanotube, and 
they predicted that an electric field in BC$_2$N nanotube induces chiral transport, analogous to a solenoid.
In Ref.~\cite{Miyamoto_1996a}, the chiral conductivity is calculated for a BC$_2$N nanotube and a carbon nanotube and is shown to be non-zero. 
Here dependence on the chirality and on the doping are not explicitly shown. 
For the BN nanotube, the dynamical current directing the chiral shift 
has been calculated. However, the dependence on the chiral angle and the electronic concentration has not been studied\cite{Kral_2000}. 
Furthermore, chiral current is investigated for metallic carbon nanotubes \cite{Tsuji_2007} and stretched carbon nanotubes \cite{Miyamoto_1996b}. Experimental  
detection of the chiral current is also proposed \cite{Miyamoto_1999}. 
The oscillation of the chiral currents in a carbon nanotube as a function of the Fermi energy 
is reported, in which the giant magnetization is expected \cite{Lambert_2008}. 
Moreover, chiral transport in 
three-dimensional chiral crystals has been also proposed recently \cite{Yoda_2015,Yoda_2018}, as well as in three-dimensional topological insulators \cite{Osumi_2021} and in two-dimensional surface and interfaces without inversion symmetry \cite{Hara_2020}.

In the present paper, we study dependence of the chiral current on the chirality of the carbon nanotubes. 
We calculate the chiral conductivity in carbon nanotubes, which is given by the
ratio between the applied electric field along the nanotube axis and the induced current around the tube.
As a function of the Fermi energy, the chiral conductivity 
is shown to 
oscillate with kinks. ThMoreover, its sign depends on $m-n$ modulo 3, where $(m,n)$ is the chirality of the nanotube. This singular behavior is interpreted in terms of the subband structure of
the nanotube, and warping of the Fermi surfaces is crucial here.
Thus, it shows that the carbon nanotube is a unique stage for chiral transport, where chiral nature can be controlled in a unique way out of the same structure. 

%

We first consider symmetry requirements to have chiral transport. 
When a current $J$ flows along the tube axis, a chiral current is induced and a magnetization $M$ along the tube will be induced. 
Thus the chiral response within the linear response 
is described as $M=CJ$ with constant $C$, and $J$ being the current along the tube axis. 
This response is called kinetic magnetoelectric effect and orbital Edelstein effect
\cite{Shalygin2012, Koretsune2012, Yoda_2015,Yoda_2018, Furukawa2017,zhong, Tsirkin2018, Furukawa2020, Rou2017, Sahin2018, Hara2020}, and 
has been studied in the three-dimensional bulk metals \cite{Yoda_2015,Yoda_2018}and in topological insulators \cite{Osumi_2021}. This effect is an orbital analog of the spin Edelstein 
effect \cite{Edelstein1990, Ivchenko1978, levitov1985nazarov}.
We will see in the following that the carbon nanotubes are special in that it contains various structures with various chiralities, based on the same two-dimensional graphene sheet.
Then we can see that 
the system should simultaneously break three types of symmetries: (i) inversion symmetry, (ii) mirror symmetry with respect to the plane 
perpendicular to the tube axis, and (iii) mirror symmetry with respect to the plane 
including 
the tube axis,  in order to realize chiral transport
induced by a current along the tube. 
The response coefficient $C$ vanishes when the system is invariant under either of these operations. 
In the three types of carbon nanotubes, i.e. armchair, zigzag and chiral ones, the symmetries 
are different as summarized in Table\ \ref{tbl:sym}. We see that only chiral nanotubes break all the three types of symmetries. in the same way as in classical solenoids. 
Thus, chiral carbon nanotubes are candidates for a chiral transport, similar to a classical solenoid. 
We note that in some three-dimensional chiral materials, similar response has been discussed. In carbon nanotubes, we will see that based on a same bulk band structure of graphene,
one can have varieties of chiral structures, and we can systematically study various chiral effects.
\begin{table}
  \begin{tabular}{|c |c|c|c |}
    \hline
       &(i) & (ii) & (iii) \\
    \hline
    armchair&$\checkmark$&$\checkmark$&$\checkmark$\\
    zigzag&$\checkmark$&$\checkmark$&$\checkmark$\\
    chiral& & &\\
    \hline
    \end{tabular}
\caption{Symmetries of three types of the carbon nanotubes. (i) inversion symmetry,
(ii) mirror symmetry w.r.t. the plane including the tube axis, and (iii) mirror symmetry w.r.t the plane perpendicular to the tube axis. }
   \label{tbl:sym}
\end{table}
 
%

\begin{figure}[htb]
\includegraphics[width=7.5cm, clip]{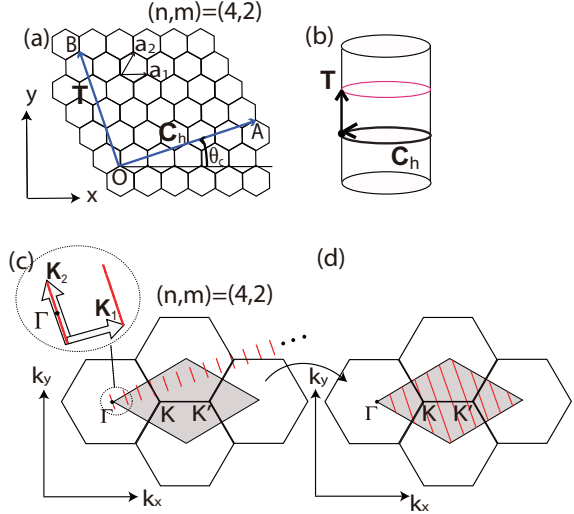} 
  \caption{Structure of carbon nanotubes and their wavevectors.  
  (a) Chiral vector ${\bf C}_h$ and the translational vector ${\rm T}$ for 
  a carbon nanotube. (a) illustrates the case with the chirality (4,2). 
(b) The vectors ${\bf C}_h$ and ${\bf T}$ in the nanotube.  (c) First Brillouin-zone for the (4,2) carbon nanotube is represented in blue. 
  The gray hexagon represents the first Brillouin zone in graphene. The wavevector along the ${\bf K}_2$ 
  is quantized, which is shown as a set of short segments shown in red. For convenience, we reexpress the
  allowed wavevectors in (d) as a set of parallel segments inside the Brillouin zone of the graphene.}
  \label{fg:nanostructure}
\end{figure}

We here explain the chirality of the the carbon nanotube. 
Figures\ \ref{fg:nanostructure}(a)(b)  show the chiral vector ${\bf C}_h$ and the translational 
vector ${\bf T}$ for the chirality $(4,2)$. 
The chiral vector ${\bf C}_h$ is a translation vector in graphene, which constitutes 
the perimeter of the nanotube,  
The chiral vector ${\bf C}_h$ is written as ${\bf C}_h=\overrightarrow{\rm OA}=n{\bf a}_1+ m{\bf a}_2$, 
with integers $m$ and $n$, where ${\bf a}_1=a(1,0)$ and ${\bf a}_2=a(1/2,\sqrt{3}/2)$ are translation vectors in graphene with 
lattice constant $a=0.246 {\rm [nm]}$ \cite{Saito_1998,Dresselhaus_1996,Ajiki_1993}. 
Chirality of the carbon nanotube is then labeled by index $(n,m)$.
The vector ${\bf T}$, which is a primitive translation vector of the nanotube, 
is expressed as

$
{\bf T}=\overrightarrow{\rm OB}=t_1{\bf a}_1+t_2{\bf a}_2
$ with $t_1=-(2m+n)/d_R$, $t_2=(2n+m)/d_R$ 
and $d_R$ is the least common multiple of $2m+n$ and $2n+m$. 
The number of hexagon $N$, contained in unit cell described by the rectangle spanned by 
${\bf C}_h$ and ${\bf T}$
in Fig.\ \ref{fg:nanostructure}(a) is given by $N=2(n^2+m^2+nm)/d_R$. 
The chiral angle $\theta _c$, which is an angle between ${\bf C}_h$ and ${\bf a}_1$, is given as 
$\theta_c=\tan^{-1}[\sqrt{3}m/(2n+m)]$. The zigzag, armchair and chiral carbon nanotubes 
can be distinguished by the chiral angle. The chiral angle is given by $\theta_c=0, \pi/3$, 
$\theta_c=\pi/6$ and $\theta_c\ (\neq 0, \pi/3, \pi/6)$, for zigzag, armchair and chiral 
carbon nanotubes, respectively. 

In the carbon nanotube, the wave number ${\bf k}$ along the circumferential direction 
is quantized \cite{Saito_1998,Dresselhaus_1996,Ajiki_1993}. 
Namely, the wave vector in the carbon nanotube is expaned in terms of ${\bf K}_1$ and ${\bf K}_2$ as
\begin{equation}
{\bf k}=k_1{\bf K}_1+ k_2 \frac{\bf K_2}{|\bf K_2|}, 
\label{eq:wavevector}
\end{equation}
with $k_1$ quantized as
\begin{equation}
k_1=0,1,2,\cdots ,N-1, -\pi/|{\bf T}|\leq k_2<\pi/|{\bf T}|. 
\end{equation}
${\bf K}_1$ and ${\bf K}_2$ are the reciprocal vectors of the carbon nanotube 
corresponding to ${\bf T}$ and ${\bf C}_h$, respectively,
satisfying
\begin{equation}
{\bf C}_h\cdot{\bf K}_1=2\pi, {\bf C}_h\cdot{\bf K}_2=0, {\bf T}\cdot{\bf K}_1=0, {\bf T}\cdot{\bf K}_2=2\pi.
\end{equation}
Namely,  
${\bf K}_1$ and ${\bf K}_2$ are given by ${\bf K}_1=(-t_2{\bf b}_1+t_1{\bf b}_2)/N$ and 
${\bf K}_2=(m{\bf b}_1-n{\bf b}_2)/N$, with ${\bf b}_1=2\pi(1,1//{\sqrt 3})$ and 
${\bf b}_2=(0,4\pi/{\sqrt 3})$. 
As an example, the first Brillouin zone of the carbon nanotube for the chirality $(4,2)$ is shown
in blue
in Fig.\ \ref{fg:nanostructure}(c). 

The energy dispersion in the carbon nanotube can be described from that in graphene. 
Namely, the energy band in the carbon naotube is given by that in graphene with the wave vector quantized along the ${\bf K}_1$ direction. 
In this paper, we use the tight-binding model with a nearest-neighbor hopping \cite{Miyamoto_1994,Hamada_1992,Saito_1992a,Saito_1992b},
\begin{equation}
H=-\sum_{\langle i,j\rangle}t a_i^{\dagger} a_j, 
\end{equation} 
where $t$ is the hopping between the sites $i$ and $j$, and 
$\langle\cdots\rangle$ means a summation over nearest-neighbor sites, 
with hopping integral 
$t\sim 3.0 {\rm eV}$ for the tight-binding model. 
The energy dispersion in this system is given by 
\begin{equation}
E_{\bf k}^{\pm}\!=\pm t\sqrt{1+4\cos^2 \frac{ak_x}{2}+
4\cos \frac{ak_x}{2}\cos \frac{ak_y}{2}} 
\end{equation}
Here, we note that the energy dispersion is exactly the
same as that in graphene, except the fact that the wave number is discretized in the carbon nanotube. 
The Fermi energy is set at $E_F=0$. It is known that 
the carbon nanotube is metallic when the wave vector crosses the Dirac points at 
$K$ and $K'$ points in graphene which corresponds to $(n-m)\equiv 0\ {\rm (mod\ 3)}$, and semiconducting for the other chiralities \cite{Ando_1998}. 
 
\section{Results and discussion}
We calculate the chiral conductivity when an electronic field is applied along the axis. 
With the Boltzmann transport equation, the conductivity is given by
\begin{equation}
\sigma_{ij}=-e^2\sum_{\bf k}^{\rm BZ}v_iv_j \frac{\partial f(E_{\bf k})}{\partial E_k}\tau,\ 
i,j=x,y,
\label{eq:Bol1}
\end{equation}
where $f(E_{\bf k})$ is the Fermi distribution function, and the summation is taken over the allowed wavevectors within the Brillouin zone (BZ), ${\bf v}=(v_x,v_y)$ is the velocity of an electron, and $\tau$ is the relaxation time assumed to be constant.
By using  $v_i=\hbar^{-1}dE_{\bf k}/dk_i$ and partial integration, at zero temperature, we obtain 
\begin{equation}
\sigma_{ij}=\frac{e^2\tau}{\hbar^2}\sum_{k_1}\int  dk_2\frac{\partial ^2E_{\bf k}^-}{\partial k_i\partial k_j}, 
\label{eq:Bol}
\end{equation}
which is a convenient form for numerical calculation. The sum over $k_1$ in eq. (\ref{eq:Bol}) is taken over the discretized wave vector $k_1$ in eq. (\ref{eq:wavevector}). 

We can calculate the chiral conductivity $\sigma_{12}$
from 
 $\sigma_{xx}$, $\sigma_{xy}$ and $\sigma_{yy}$ in eq.~(\ref{eq:Bol}), by
rotating the coordinate axes from $x,y$ to $1,2$ bythe angle $\theta_c$. 
As a result, the chiral conductivity $\sigma_{12}$ is obtained as 
\begin{eqnarray}
\begin{pmatrix}
\sigma_{11} &\sigma_{12} \\
\sigma_{21} & \sigma_{22}
\end{pmatrix}
=
\begin{pmatrix}
\cos \theta_c & \sin\theta_c \cr -\sin \theta_c & \cos \theta_c \cr
\end{pmatrix}
\nonumber \\
\times
\begin{pmatrix}
\sigma_{xx} & \sigma_{xy} \cr \sigma_{yx}& \sigma_{yy} \cr 
\end{pmatrix}
\begin{pmatrix}
\cos \theta_c & -\sin\theta_c \cr \sin \theta_c & \cos \theta_c \cr 
\end{pmatrix}
, \label{eq:sigma1112}
\end{eqnarray}
where $1$ and $2$ represent the circumferential direction and the
 direction along the tube, 
respectively. 
We calculate the chiral conductivity $\sigma_{12}$ by varying
two parameters, the chirality $(n,m)$ and the electronic concentration $n_i$. 
Because the band dispersion depends on chirality, 
the Fermi level is numerically determined for an electronic concentration $n_i$  for each carbon nanotube. 
 
\begin{figure}[htb]
\includegraphics[width=8.5cm]{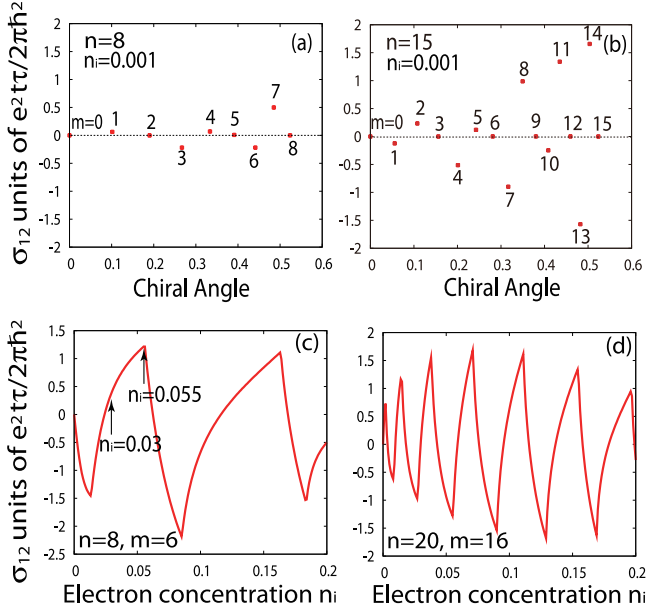}
\caption{Chiral conductivity $\sigma_{12}$. (a)(b) Chiral conductivity as a function of the chiral angle for $n_i=0.001$. 
The chirality is set as (a) $(8,m)$ with $m=0,1,\cdots,8$ and 
(b) $(15,m)$ with $m=0,1,\cdots,15$.  (c)(d) Chiral conductivity as a function of the electron concentration for (c) (8,6) and (d) (20,16) nanotubes.}
\label{fg:815ni0001}
\end{figure}

Figures\ \ref{fg:815ni0001} (a) and (b) shows the chiral conductivity $\sigma_{12}$ as a 
function of the chiral angle for the electronic concentration $n_i=0.001$, measured from the undoped value. 
We set the chirality as $(8,m)$ in (a) and $(15,m)$ in (b), where integer $m$ is changed from $0$ to $8$. 
Generally, the sign of the chiral conductivity $\sigma_{12}$ depends on $m$
through the value $(n-m)\equiv0,1,2\ {\rm (mod\ 3)}$. 
Namely, 
for $(n-m) \equiv1\ (2)\ {\rm (mod\ 3)}$, the sign of $\sigma_{12}$ is positive (negative).
For $(n-m)\equiv0\ {\rm (mod\ 3)}$ when the carbon nanotube becomes metallic,
$\sigma_{12}$ is almost zero. We note that $\sigma_{12}$ becomes zero for $m=0$ 
because the carbon nanotube is of the armchair type and chiral transport is 
prohibited by symmetry. 
Furthermore, 
$|\sigma_{12}|$ tends to become larger as the chiral angle increases, 
with an exception of $m=5,10$ for $n=15$.
\par

We found
that $\sigma_{12}$ tend to become larger as the integers $m$ and $n$ in the chirality $(m,n)$ become larger in 
general, by calculating 
also for the chiralities $(4,m)$, $(12,m)$, $(20,m)$ and $(30,m)$, although not shown here. 
This is because the contribution from the Fermi surfaces to $\sigma_{12}$ becomes larger for large chirality. The density of states on the Fermi level increases as chirality increases, with some exceptions with accidental local minima of $\sigma_{12}$ such as those for 
$m=5,10$ in Fig.\ \ref{fg:815ni0001}(b). 
We interpret these remarkable dependences of $\sigma_{12}$ on chirality in terms of the warping of the Fermi surfaces. 
For large doping, the warping effect is prominent and the contributions to  $\sigma_{12}$ do not perfectly cancel each other in metallic carbon nanotube, as we discuss later. 
 
Figures\ \ref{fg:815ni0001}(c) and (d) shows $\sigma_{12}$ as a function of electronic concentration $n_i$ 
for $0\leq n_i\leq0.2$ for the chiralities (a) $(8,6)$ and (b) $(20,16)$. 
Generally, $\sigma_{12}$ oscillates as a function of $n_i$, with sharp kinks where 
$\sigma_{12}$ changes between an increasing and a decreasing function of $n_i$.  
As $n_i$ becomes even larger and approaches 1, which is not realistic but is studied only for a theoretical interest, the chiral conductivity converges to zero, because electrons 
occupy all the bands in the model, giving no carriers in the hole picture. 

\begin{figure}[htb]
\includegraphics[width=8.5cm]{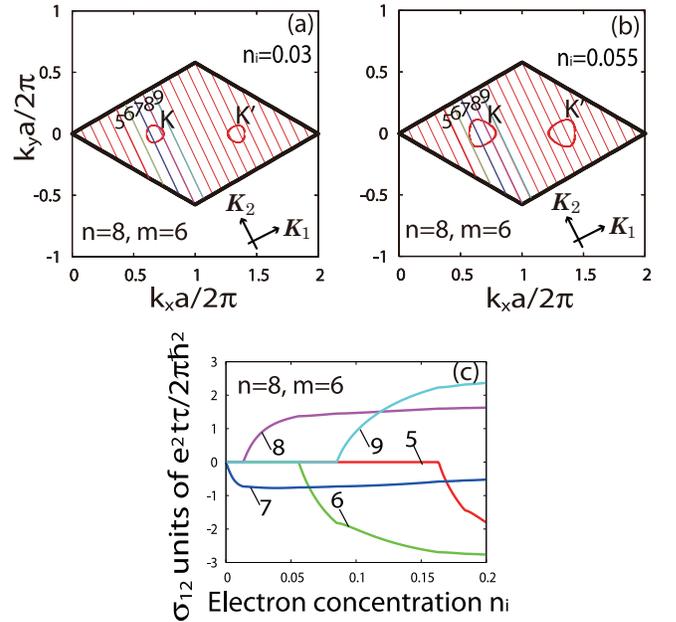}
\caption{Singular behaviors of chiral condiuctivities from subband contributions. The Fermi surface of the $(8,6)$ carbon nanotube is shown 
for (a) $n_i=0.03$ and 
(b) $n_i=0.055$ in the first Brillouin zone in graphene. The lines are the discretized wave vectors 
of carbon nanotubes. The indexes $5$-$9$ are subband indices  showing the values of $k_1$.
(c) The contribution from the $l$-th band ($l=5,6,7,8,9)$ to $\sigma_{12}$ as a function of 
the electronic concentration $n_i$ for the chirality $(8,6)$. 
Each color and number correspond to those in (a) and (b). }
\label{fg:86niband}
\end{figure}

To explain the oscillation of $\sigma_{12}$, we analyze the dependence of $\sigma_{12}$ on $n_i$ in the $(8,6)$ carbon 
nanotube for $0\leq n_i\leq0.2$ in Fig.\ \ref{fg:815ni0001} (c). 
As an example, in Fig.\ \ref{fg:86niband}(a) we show the Fermi surface at $n_i=0.03$, where $\sigma_{12}$ monotonically 
increases. 
The rhombus in the figure is the unit cell for the reciprocal space of graphene, which is equivalent to the first Brillouin zone. 
The wavevector ${\bf k}$ is discretized along the ${\bf K}_1$ direction (see eq.~(\ref{eq:wavevector})), shown as the parallel lines, giving 
rise to subband structures.  We label each subband by the value of the integer $k_1$, 
and they are shown as the indices
from $5$ to $9$ for the subbands near the Fermi level. 
There are two Fermi surfaces, near the K and the K' points for the graphene. 
At $n_i=0.03$, 
the Fermi level crosses only the $7$th and $8$th lines. 
The chiral conductivity $\sigma_{12}$ in eq.~(\ref{eq:Bol1}) is attributed to the states at the Fermi energy.
Here we focus on the Fermi surface near the K point, because the contribution from the states near 
the K point and that near the K' point are the same by the time-reversal symmetry.
The states on the Fermi level near the K point are represented by four intersection points between the 
Fermi surface and the allowed wavevectors specified by the parallel lines. 
On the other hand, 
for  $n_i=0.055$, where the chiral conductivity $\sigma_{12}$ shows a peak in 
Fig.\ \ref{fg:815ni0001} (c), the Fermi surface is illustrated in Fig.\ \ref{fg:86niband} (a). 
In addition to the four intersection points on the $7$th and the $8$th lines in Fig.\ \ref{fg:86niband} (a), 
the Fermi surface and the $6$th line come in contact with each other in this case. 
To see how each band contributes to the chiral conductivity $\sigma_{12}$, 
we show in Fig.\ \ref{fg:86niband} (c) the contributions to $\sigma_{12}$ from each 
subband 
as a function of $n_i$. 
We can see the negative contribution from the 6th band in Fig.\ \ref{fg:86niband} (c)
appears for $n_i>0.055$. Therefore,  
the sharp kink of  $\sigma_{12}$ at $n_i=0.055$ is attributed to the negative contribution of the 6th band to $\sigma_{12}$ for $n_i>0.055$.
As the electron concentration increases, the Fermi surface crosses the 
7th, 8th, 6th, 9th and 5th bands consecutively, and they contribute to the 
chiral conductivity with various signs.
This corresponds to the points of the sharp kinks in Fig.\ \ref{fg:815ni0001} (c). 
Besides, the signs of the contributions changes alternately in the order, 7th, 8th, 6th, 9th and 5th, which 
explains the oscillation of $\sigma_{12}$ as a function of $n_i$. 
 
Figure\ \ref{fg:815ni0001} (d) shows $n_i$ dependence of $\sigma_{12}$ for the chirality 
$(20,16)$. The range of the oscillation is larger, and the period of the oscillation is smaller 
than those for the chirality $(8,6)$. 
\begin{figure}[htb]
\includegraphics[width=8.5cm]{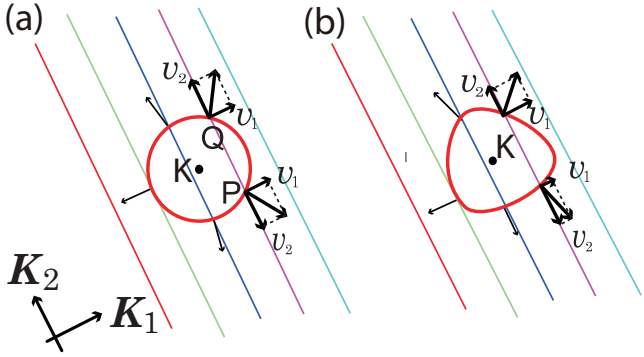}
\caption{Effect of the warping of the the Fermi surface onto the chiral conductivities. The Fermi surface (a) without and (b) with warping near the $K$ point in graphene. 
The lines describe the subbands of the carbon nanotube. 
The arrows normal to the Fermi surface denote the Fermi velocity, which can be 
decomposed into the circumferential velocity $v_1$ and the velocity along the tube $v_2$.} 
\label{fg:image}
\end{figure}
 
To understand the reason for the singular behavior of $\sigma_{12}$, we first assume the 
band structure around the $K$ point to be a perfect Dirac cone 
with perfect linear dispersion $E_{\bf k}=\gamma |{\bf k}-{\bf k}_0|$ where $\gamma$ is a constant, 
and ${\bf k}_0$ is either the K or K' point. 
Then $\sigma_{12}$ is completely zero for any chiralities and the electronic 
concentrations. Since the linear dispersion $E_{\bf k}=\gamma |{\bf k}-{\bf k}_0|$ makes the Fermi surface to be a circle without warping, there are always a pair of Fermi points in the same subband, 
where the values of $v_1$ are the same while the signs of $v_2$ are opposite (the points P and Q in  Fig.\ \ref{fg:image} (a)). Their contributions to $\sigma_{12}$ in eq.~(\ref{eq:Bol1}) 
completely cancels each other because of symmetry. 
In reality, 
the Fermi surfaces in carbon nanotubes are distorted from a circle and have a warping (Fig.\ \ref{fg:image} (b)). 
For the Fermi surface with the warping, the chiral conductivity has a nonzero value 
in general, because the contributions from the Fermi points do not necessarily 
cancel each other. 

For $n-m\equiv 0\ ({\rm mod}\ 3)$, one of the subbands crosses the Dirac point in graphene.  
In this case, the contributions cancel each other almost completely for the low electronic concentration such as $n_i=0.001$, because the Fermi surface is 
almost symmetric with respect to the subband line going across the Dirac point. 
However, the contribution does not vanish completely each other for a higher electronic 
concentration, because of the large warping of the Fermi surface.
 
\begin{figure}[htb]
\includegraphics[width=8.5cm]{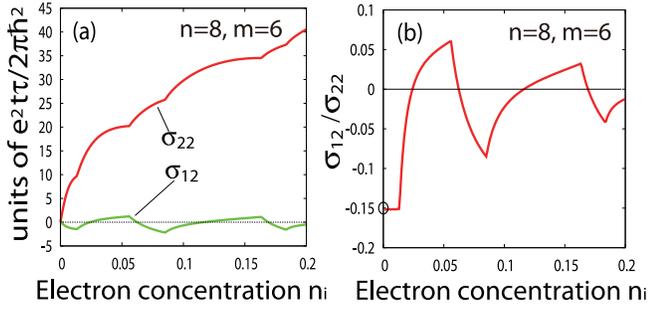}
\caption{The longitudinal conductivity $\sigma_{22}$ and the chiral conductivity $\sigma_{12}$. (a) $\sigma_{12}$ and $\sigma_{22}$ as functions of $n_i$ in the $(8,6)$ nanotube. 
(b) $\sigma_{12}/\sigma_{22}$ as a function of $n_i$ in the $(8,6)$ nanotube.  } 
\label{fg:exp}
\end{figure}

We now estimate the value of the current flowing 
along the circumferential direction. 
When the electric field is applied along the axis, the current per unit length along the circumferential direction 
$j_1$ is given by 
\begin{equation}
j_1=\frac{\sigma_{12}}{\sigma_{22}}\frac{I_2}{2\pi R}, 
\label{eq:j2}
\end{equation}
where $I_2$ is the current along the nanotube, and $R$ is the radius of 
the carbon nanotube. We note that $\sigma_{22}$ 
is the longitudinal conductivity along the tube axis, calculated from 
eq.~(\ref{eq:sigma1112}).
In Fig.\ \ref{fg:exp}(a), we show the comparison of $\sigma_{22}$ and $\sigma_{12}$ 
in the $(8,6)$ carbon nanotube, where the diameter is $R=0.47{\rm nm}$. 
$\sigma_{22}$ increases with some oscillations as $n_i$ increases. 
 $\sigma_{22}$ shows a moderate kink at the electronic concentrations 
where $\sigma_{12}$ shows sharp peaks and dips. 
This is becasue new subbands begin to contribute to $\sigma_{12}$ and$\sigma_{22}$, as we increase the concentration, 
as we have seen in this paper.
In Fig.\ \ref{fg:exp}(b), we show the ratio between $\sigma_{22}$ and $\sigma_{12}$, which 
represents a dimensionless figure of merit for the chiral transport.
We note that the ratio becomes largest at a very low concentration, and that it 
oscillates, converging to zero for larger $n_i$ because $\sigma_{22}$ 
increases rapidly as the Fermi level increases. 
The positions of the peaks of the oscillation correspond to that of $\sigma_{12}$ and $\sigma_{22}$. 
For a very low concentration region, the ratio takes a value $\sim-0.15$ in this case, and becomes zero at $n_i=0$. 
For the $(8,6)$ nanotube, the maximum current which can be applied to 
the carbon nanotube is $I_1^{\rm max}\times S=6.6\times 10^{-8}{\rm A} $ \cite{Dekker_1999}, 
with the cross section of the $(8,6)$ nanotube $S=6.6\times10^{-17}\rm cm^2$. 
For $\sigma_{12}/\sigma_{22}\sim -1.5$, $j_1$ is $j_1=-3.5{\rm A} /{\rm cm}$ from eq.\ (\ref{eq:j2}), corresponding to 4.4 gauss.
The (orbital) magnetization per unit length of the nanotube is estimated as 
$M=j_2S=-0.23\times 10^{-15}{\rm A}\cdot {\rm cm}$. 

\section{Concluding remarks}
In conclusion, we studied chiral transport in carbon nanotubes. 
We calculated the chiral conductivity by using the tight-binding model 
and the Boltzmann transport equation for the electric field applied along the axis 
of the carbon nanotube. 
We found that the chiral conductivity $\sigma_{12}$ is nonzero for semiconducting chiral carbon nanotubes with doping. 
For a constant electronic concentration, its sign alternately changes 
as the chiral angle increases, and it becomes positive, negative and zero when for the cases with $n-m\equiv1,2,0 {\rm (mod 3)}$, respectively. 
As a function of the electron concentration, $\sigma_{12}$ oscillates.  
This is attributed to the different signs of the contributions from the individual subbands to $\sigma_{12}$. 
As the electronic concentration increases, the Fermi surface crosses the subbands with positive and negative contributions to $\sigma_{12}$, giving rise to a peculiar oscillatory dependence on the concentration. 
Finally, we have estimated the value of chiral current and magnetization induced by the chiral current.

This kind of chiral transport has been also studied in bulk metals and topological insulators in three dimensions. For example, in tellurium, this chiral transport has been evaluated by {\it ab initio} calculation \cite{Tsirkin2018} and  a related experiment on current-induced magnetization has been performed \cite{Furukawa2017}. Tellurium hss left-handed and right-handed crystals, which consist of one-dimensional chains in 
shapes of helices, and one can see its analogy with a classical 
solenoid easily. On the other hand, in chiral nanotubes, its chirality is not fixed but can switch between 
both directions by changing an electron concentration. 

This effect induces an orbital magnetization in carbon nanotubes, and through the spin-orbit
coupling it might also lead to spin magnetization, which might be one route to induce the  chirality-induced spin selectivity (CISS) effect\cite{Naaman2012, Gohler2011, Kettner2018, Bloom2016, Dor2014}.

\vspace{2mm}


{\bf Notes}\\
The authors declare no competing financial interest.

\begin{acknowledgement}
This work was supported by JSPS KAKENHI Grants No. JP30578216, No.~23740236,  No.26287062, and No. 18H03678, 
by the JSPS-EPSRC Core-to-Core program ''Oxide Superspin",
by
MEXT KAKENHI Grants No. 25103709, and No.26103006, 
and by Elements strategy Initiative to Form Core Research Center (TIES), from MEXT Grant Number JP-MXP0112101001.
\end{acknowledgement}

\providecommand{\latin}[1]{#1}
\makeatletter
\providecommand{\doi}
  {\begingroup\let\do\@makeother\dospecials
  \catcode`\{=1 \catcode`\}=2 \doi@aux}
\providecommand{\doi@aux}[1]{\endgroup\texttt{#1}}
\makeatother
\providecommand*\mcitethebibliography{\thebibliography}
\csname @ifundefined\endcsname{endmcitethebibliography}
  {\let\endmcitethebibliography\endthebibliography}{}


\end{document}